\documentstyle[aps]{revtex}
\begin{document}
\draft
\title{
Multifractal Analysis of the Dynamical Heterogeneity 
in a Two-Dimensional Supercooled Liquid 
\cite{N}}
\author{
W. Sakikawa and O. Narikiyo}
\address{
Department of Physics, 
Kyushu University, 
Fukuoka 810-8560, 
Japan}
\date{
May, 2002}
\maketitle
\begin{abstract}
The dynamical heterogeneity in supercooled liquids 
measured by a molecular dynamics simulation 
has been quantified on the basis of the multifractal formalism. 
The singularity spectrum becomes broader as the glass transition 
is approached. 
This behavior is similar to that observed 
in the case of the Anderson-localization transition. 
\end{abstract}
\vskip 18pt
  The dynamical heterogeneity in supercooled liquids 
has been studied intensively in recent years. 
  It can be quantified by introducing the concept of bond-breakage 
and the distribution of the broken bonds near the glass-transition 
is found to be similar to that of the spin state of the Ising model 
near the critical point.\cite{YO} 
  The bond-breakage corresponds to the collective jump motion 
of particles among cages. 

  On the other hand, we have analyzed the dynamical heterogeneity 
of the critical spin state of the Ising model 
on the basis of the multifractal formalism.\cite{SN} 
  In this Short Note we try to combine above two approaches 
and quantify the dynamical heterogeneity 
observed in a molecular dynamics simulation of a supercooled liquid 
in terms of multifractality. 
  The so-called cooperatively rearranging regions (CRR) 
can be identified with weakly bonded regions with various sizes 
which form multifractal structure. 

  We use the same model as in Ref.\ 1,  
the binary mixture of soft core particles. 
  Two species of particles, 1 and 2, 
with the size $\sigma_{1}$ and $\sigma_{2}$, 
and the mass $m_{1}$ and $m_{2}$ are considered. 
  The interaction potential is given by 
\begin{equation}
v_{\alpha\beta}(r)
 = \epsilon \cdot ({\sigma_{\alpha\beta} \over r})^{12}, 
\end{equation}
where $r$ is the distance between two particles and 
$\sigma_{\alpha\beta}=(\sigma_{\alpha}+\sigma_{\beta})/2$ 
with $\alpha,\beta = 1,2$. 

  The molecular dynamics simulation is done 
using the second order symplectic integrator\cite{TBM} 
and the microcanonical ensemble. 
  The temperatures in the following are determined 
by the kinetic energy. 

  The bond is introduced as in Ref.\ 1 when 
\begin{equation}
r_{ij}(t_{0}) \leq 1.1\sigma_{\alpha\beta},  
\end{equation}
where $r_{ij}(t_{0})$ is the distance between 
$i$-th particle of species $\alpha$ and 
$j$-th particle of species $\beta$ at time $t_{0}$. 
  The position of the bond is represented by the midpoint 
of the two particles at this time. 
  The bonds are unambiguously defined\cite{YO} 
owing to the sharpness of the first peak 
of their pair correlation function. 
  The bond is regarded as surviving at time $t_{0}+t$ when 
\begin{equation}
r_{ij}(t_{0}+t) \leq 1.6\sigma_{\alpha\beta}.  
\end{equation}
  The number of surviving bonds $N_{\rm suv}(t_{0}+t)$ is fitted by 
\begin{equation}
N_{\rm suv}(t_{0}+t) = N_{\rm suv}(t_{0})\exp(-t/\tau_{\rm b}),  
\end{equation}
introducing the life-time of bonds $\tau_{\rm b}$. 
  In the following the broken bonds are determined for $t=0.05\tau_{\rm b}$. 

  In our simulation the followings are fixed.  
  Two species of particles have the same particle number, 
$N_{1} = N_{2} = 80000$. 
  The size and mass ratios are chosen as 
$\sigma_{2}/\sigma_{1} = 1.4$ and $m_{2}/m_{1} = 2$ respectively. 
  The time step of the integration $\Delta t$ is chosen 
as $\Delta t = 0.005\tau_{1}$ 
with $\tau_{1} = (m_{1} \sigma_{1}^{2}/\epsilon)^{1/2}$. 
  The potential is truncated at $r = 4.5\sigma_{1}$. 
  The particle density is chosen as $n = 0.8/\sigma_{1}^{2}$. 
  The simulations were performed on our personal computer 
with Pentium-4 CPU. 

  In Fig.\ 1 the snapshot of the broken bonds is shown 
where the position of the bond is defined at time $t_{0}$ 
as mentioned in the above. 
  In Figs.\ 1(a) and 1(b) the temperatures are chosen as 
$T = 2.54$ and $T = 0.859$ respectively 
in the unit of $\epsilon/k_{\rm B}$. 
  These two states are realized starting from an initial state 
with the temperature 9.69 and cooled in a stepwise manner as in Ref.\ 1. 
  In the stepwise cooling process 
the temperatures in the first, second, third and fourth stages are 
9.69, 2.54, 1.43 and 0.859 respectively. 
  For the former state the temperature was changed from 9.69 to 2.54 
after spending 5000$\Delta t$ at the first stage 
and the broken bonds are observed at the second stage 
with $t_{0}=5000\Delta t$. 
  For the latter state the temperature was changed successively 
from 9.69, 2.54, 1.43 to 0.859 after spending 5000$\Delta t$ 
at the first, second and third stages each 
and the broken bonds are observed at the fourth stage 
with $t_{0}=60000\Delta t$. 
  The times $t_{0}$ and $t$ are measured from the instance 
of the temperature change to the stage of the observation. 
  The life-times of bonds were measured to be 
$\tau_{\rm b}=4.09 \times 10^{3}\Delta t$ for $T=2.54$ 
and $\tau_{\rm b}=2.43 \times 10^{5}\Delta t$ for $T=0.859$. 

  Qualitatively the snapshot shows a multifractal behavior and 
the degree of intermittency increases as the temperature is decreased. 
  In the following we quantify this observation 
on the basis of the multifractal formalism. 

  The multifractality is quantified by the singularity spectrum, 
$f(\alpha)$, and the spectrum is obtained by the following formulae,\cite{S} 
\begin{equation}
\alpha(q) = \sum_{k=1}^{N_L} \mu_{k}(q,l) \ln \mu_{k}(1,l) / \ln (l/L),  
\end{equation}
and 
\begin{equation}
f(\alpha(q)) = \sum_{k=1}^{N_L} \mu_{k}(q,l) \ln \mu_{k}(q,l) / \ln (l/L),  
\end{equation}
with 
\begin{equation}
\mu_{k}(q,l) = \{\mu_{k}(l)\}^{q} 
             / \sum_{k'=1}^{N_L} \{\mu_{k'}(l)\}^{q}. 
\end{equation}
  Here the measure, $\mu_{k}(l)$, is defined as the probability 
to find a broken bond in the $k$-th square box 
of the linear dimension $l$ 
in a snapshot of broken bonds. 
  The total number of the boxes $N_L = (L/l)\times(L/l)$ 
where $L$ is the linear dimension of the simulation region 
with periodic boundary condition. 

  In Fig.\ 2 the singurarity spectra 
for $T = 2.54$ and $T = 0.859$ are shown. 
  The dots represent the data set $(\alpha(q), f(\alpha(q)))$ determined 
by the above formulae and should be tangential 
to $f(\alpha)=2$ and $f(\alpha)=\alpha$. 
  The maximum $\alpha_{\rm max}$ and minimum $\alpha_{\rm min}$ 
are determined by the intercept of the spectrum extrapolated 
to the $\alpha$-axis 
and correspond to the singularity exponent for the box 
with the lowest density of broken bonds 
and the one with the highest density respectively. 

  From the data in Fig.\ 2 
we can quantify the degree of intermittency. 
  The singularity spectrum becomes broader 
as the temperature is decreased approaching the glass transition. 
  At the same time $\alpha_{\rm max}$ becomes larger 
and $\alpha_{\rm min}$ becomes smaller. 
  Namely $\alpha_{\rm max}=2.74$ and $\alpha_{\rm min}=1.66$ for $T=2.54$ 
and $\alpha_{\rm max}=2.97$ and $\alpha_{\rm min}=1.57$ for $T=0.859$. 
  This finding is similar to 
that reported in the case of the Anderson localization\cite{S} 
where the singularity spectrum becomes broader 
as the strength of the impurity potential is increased 
and the localization transition is approached. 
  In both cases 
the degree of intermittency corresponds to that of localization. 

  In conclusion 
we have quantified the dynamical heterogeneity 
in a model supercooled liquid 
measured by a molecular dynamics simulation 
on the basis of the multifractal formalism. 
  In supercooled liquids 
various sizes of CRR are formed and constitute multifractal structures. 
  The degree of intermittency increases approaching the glass transition. 
  Since our simulation parameters are still far from 
those for the glass transition, 
we need further study 
to elucidate the nature of the glass transition. 

  The authors are grateful to Jun Matsui 
for enlightening discussions on the molecular dynamics simulation. 
  This work was supported in part 
by a Grand-in-Aid for Scientific Research 
from the Ministry of Education, Culture, Sports, Science 
and Technology of Japan. 


\vskip 15pt
\begin{figure}
\caption{
The snapshot of the broken bonds 
for (a) $T=2.54$ and (b) $T=0.859$. }
\label{fig:1}
\end{figure}

\vskip -15pt

\begin{figure}
\caption{
The singularity spectrum 
for $T=2.54$ and $T=0.859$ with $L/l=30$. }
\label{fig:2}
\end{figure}

\begin{references}

\bibitem{N}
This paper has been submitted to the Short Note section 
of J. Phys. Soc. Jpn.

\bibitem{YO}
R.\ Yamamoto and A.\ Onuki: 
J. Phys. Soc. Jpn. {\bf 66} (1997) 2545. 

\bibitem{SN}
W.\ Sakikawa and O.\ Narikiyo: 
J. Phys. Soc. Jpn. {\bf 71} (2002) 1200. 

\bibitem{TBM}
M.\ Tuckerman, B.\ J.\ Berne and G.\ J.\ Martyna: 
J. Chem. Phys. {\bf 97} (1992) 1990. 

\bibitem{S}
M.\ Schreiber: 
{\it Computational Physics}, ed. K.\ H\ Hoffmann and M.\ Schreiber 
(Springer-Verlag, Berlin, 1996) p.147. 

\end{references}
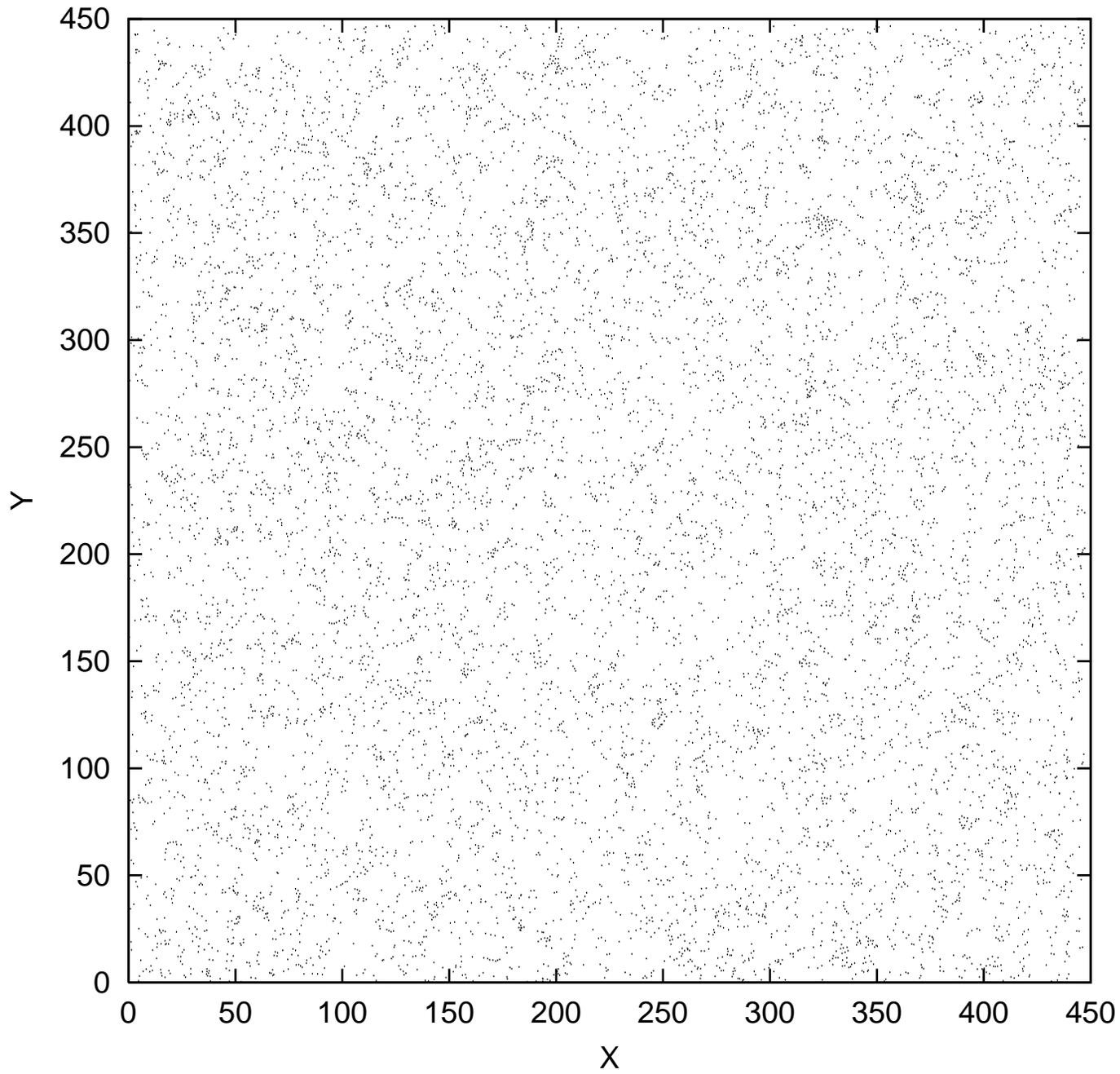
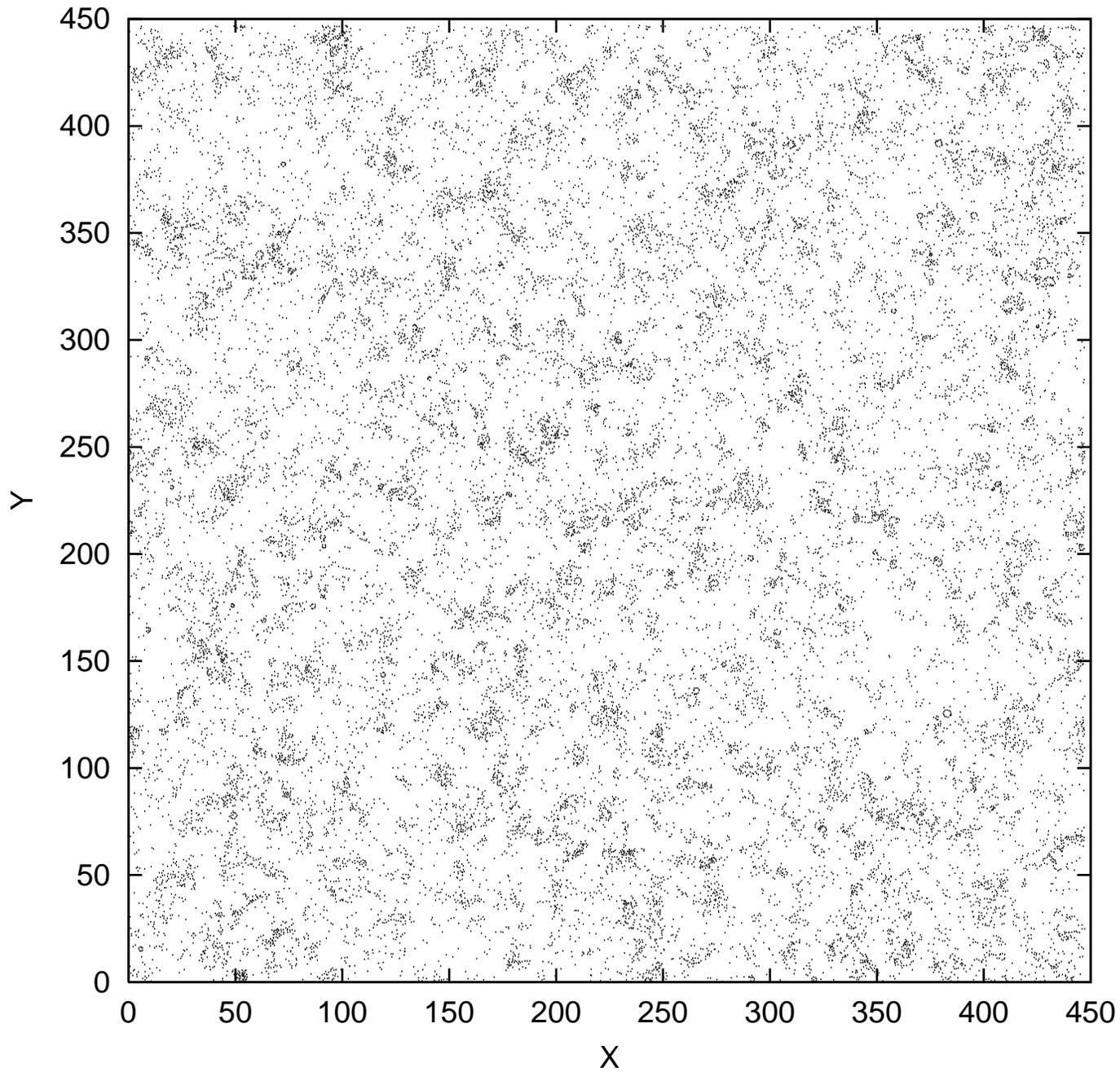
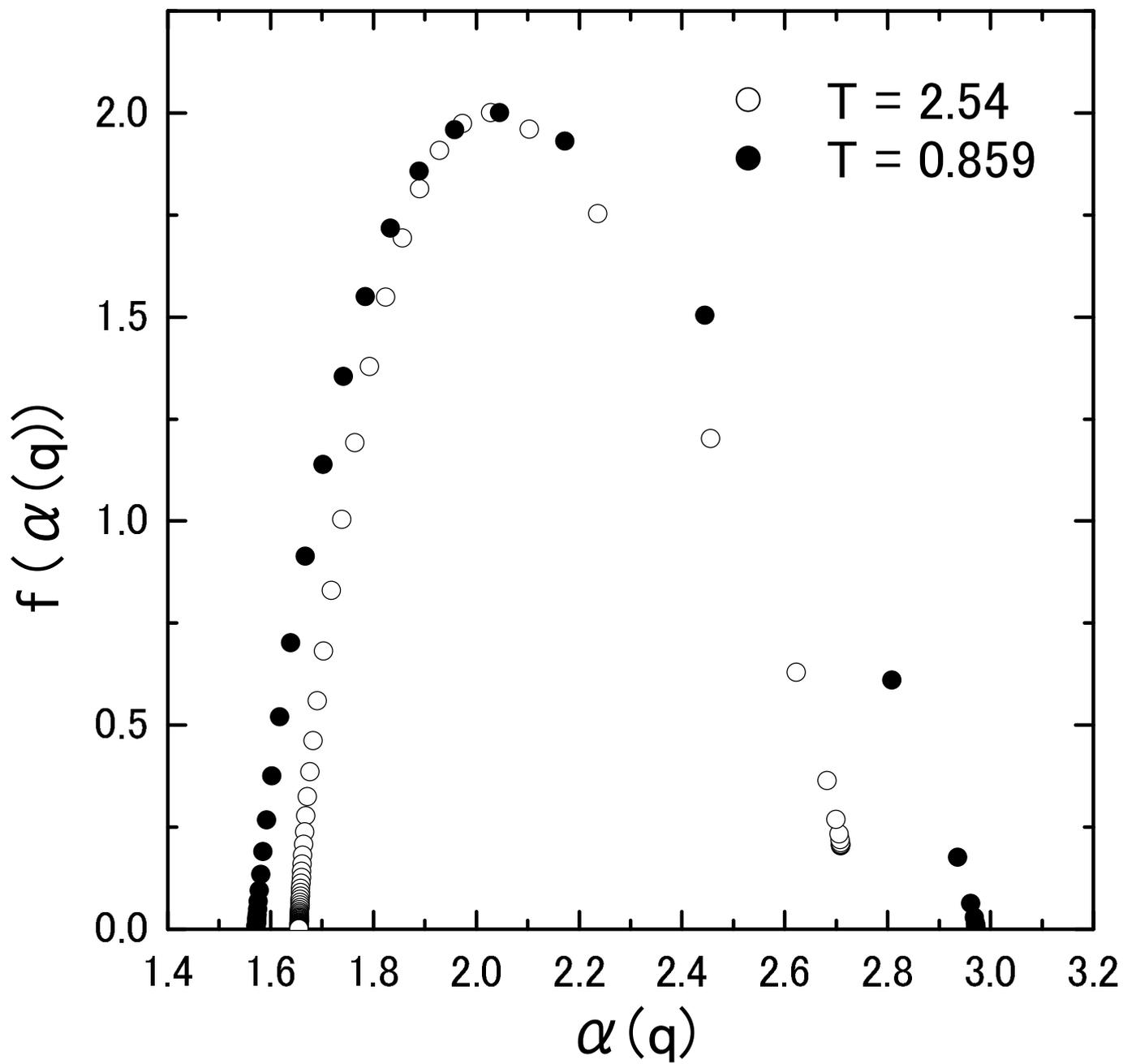
\end{document}